\documentclass[a4wide,12pt]{article}
\usepackage{amssymb,amsmath,often_used_commands}
\usepackage{epsfig}
\usepackage{a4}
\usepackage{parskip}
\usepackage{color}
\usepackage{relsize}
\usepackage{multirow}
\newcommand{\sect}[1]{\setcounter{equation}{0}\section{#1}}

\def\N{{\mathcal N}}
\def\L{{\mathcal L}}

\def\O{{\mathcal O}}
\def\L{{\mathcal L}}

\def\a{\alpha}
\def\b{\beta}

\def\m{\mu}
\def\n{\nu}

\def\f{\phi}
\def\vf{\varphi}
\def\e{\eta}

\def\l{\lambda}

\def\sinh{\mathrm{sinh}}
\def\cosh{\mathrm{cosh}}

\def\coth{\mathrm{coth}}
\def\p{\partial}
\def\rb{\right}
\def\lb{\left}
\def\axs{AdS_5\times S^5}

\newcommand{\eq}[1]{\begin{equation} #1 \end{equation}}
\newcommand{\eqs}[1]{\begin{equation*} #1 \end{equation*}}
\newcommand{\al}[1]{\begin{align} #1 \end{align}}
\newcommand{\als}[1]{\begin{align*} #1 \end{align*}}
\newcommand{\ml}[1]{\begin{multline} #1 \end{multline}}

\begin{document}

\begin{titlepage}

\begin{flushright}{\small ESI--1983\\ \small TUW--07--16}\end{flushright}

\vspace*{1.5cm}

\begin{center}
{\Large \textbf{A note on the Near Flat Limit for strings in the Maldacena-Nunez background}}

\vspace*{1cm}
 M. Kreuzer${}^{\star}$\footnote{e-mail:kreuzer@hep.itp.tuwien.ac.at}, 
C. Mayrhofer ${}^{\star}$\footnote{e-mail:cmayrh@hep.itp.tuwien.ac.at}
and 
R.C. Rashkov${}^{\dagger\star}$\footnote{e-mail:rash@hep.itp.tuwien.ac.at; 
on leave from Dept of Physics, Sofia University, Bulgaria.}

\ \\

${}^{\star}$ \textit{Institute for Theoretical Physics, Vienna
University of Technology,\\
Wiedner Hauptstr. 8-10, 1040 Vienna, Austria}

\ \\

${}^{\dagger}$ \textit{
Erwin Schr\"{o}dinger International Institute
for Mathematical Physics, \\
Boltzmanngasse 9, A-1090 Wien, Austria
}
\end{center}

\vspace*{.8cm}

\baselineskip=18pt

\bigskip
\bigskip
\bigskip

\begin{abstract}
Recently Maldacena and Swanson suggested a new limit of string theory on the $AdS_5\times S^5$
background, the so called near flat space limit. The resulting reduced theory interpolates between
the  pp-wave limit and giant magnon type string solutions. It was shown that the reduced model 
possess many features of the original theory. On the other hand, theories with less supersymmetry
are of great importance for the string/gauge theory correspondence. In this paper we study the near 
flat limit reduction of string theory on the Maldacena-Nunez background, which is dual to $\N=1$
Yang-Mills theory. The reduced model interpolates between the pp-wave limit and a certain magnon type subsector
of the theory. The similarity of the structures of the  reduced model obtained here and that by
Maldacena and Swanson indicates the possibility of existence of integrable subsectors of strings on the Maldacena-Nunez 
background. 
\end{abstract}

\end{titlepage}

\sect{Introduction}

The idea for the correspondence between the large N limit of gauge
theories and strings emerged over thirty years ago
\cite{'tHooft:1973jz} and since then many attempts for explicit realization
have been made. One of the most promissing so far was
provided by Maldacena who conjectured the AdS/CFT correspondence
\cite{holography}. Since then this became a topical area
and many fascinating discoveries were made over the last decade.
One of the predictions of the correspondence
is the conjectured one-to-one map between the spectrum of  string theory on
$\axs$ and the spectrum of anomalous dimensions of gauge invariant
operators in the  $\N=4$ Supersymmetric Yang-Mills (SYM)
theory.

The vast majority of papers were on qualitative and quantitative
description of $\mathcal{N}=4$ SYM theory with $SU(N)$ gauge group
by making use of the string sigma model on $AdS_{5}\times S^{5}$. This special case
is remarkable with its classical integrability \cite{BPR} and it is hardly believed that the
integrability is preserved at quantum level as well.
Thus, one can put forwards the important issue of derivation of the S-matrix and 
study the main features of the string model and its dual.
This issue is important because one of the predictions of the correspondence is
the equivalence between the spectrum of string theory on
$\axs$ and the spectrum of anomalous dimensions of gauge invariant
operators. 
There has been a good deal of success recently in comparing the energies  of
semiclassical strings and the anomalous dimensions of the gauge theory operators. 
However, on one side string theory on $AdS_{5}\times S^{5}$ is highly
non-linear and on another the tools at hand to study gauge theory are limited mainly to
perturbation theory. 
Having in mind the duality, i.e. complementarity of the
validity regions of the perturbation theories, the proof of the conjecture becomes a 
real challenge. Luckily, there exists a wide region where one can trust quasi-classical considerations
and compare the corresponding spectra on both sides. Integrability provides powerful tools for
calculating the key features on each side of the correspondence.

There are several approaches using simplifications of the non-linear system we are dealing with. 
An important proposal by Berenstein, Maldacena and Nastase \cite{bmn}
made the first step beyond the supergravity approximation in the AdS/CFT 
correspondence. They showed how certain operators in SYM theory 
can be related to string theory in the pp-wave limit in a certain way. 
Namely, it was suggested that string theory in pp-wave background 
gives good approximation of a certain class of "nearly" chiral operators
and vice versa. Since string theory in the pp-wave 
background is exactly solvable, a lot of information for the AdS/CFT correspondence was 
extracted\footnote{For extensive review see for instance \cite{JPlefka:2004}
\cite{DSadri:2004}.}.

Superstring theory on $\axs$ is integrable at the classical level and there exist strong indications 
 that it happens also at the quantum  level.  Nevertheless, it is a highly  non-trivial task to 
find the exact S-matrix of the theory and to study the duality in its full extent. 
The difficulties are not only of technical nature but also conceptual. Moreover, one must 
keep considerations such that one can check the results on both sides of the correspondence. 
The latter assumes certain limits \cite{GKP},
allowing reliability of the semiclassical results on both sides. In this case the techniques of 
integrable systems have become useful to study the AdS/CFT correspondence in detail. 
For instance, the remarkable observation by Minahan and Zarembo
 \cite{Minahan:2002ve} made it possible to relate gauge theory side of the correspondence to some 
integrable spin chain 
(at least to leading orders). Thus, the question of relating string theory side to some
spin chain and then compare to the gauge theory side becomes very important. 
If so, the integrability of spin chains would provide a powerful tool for investigation
of AdS/CFT correspondence. 
Discussing the integrability properties of the superstring on $\axs$, one should point out 
several key developments leading to the current understanding. One of them is
the gauge theory S-matrix introduced in \cite{staud:0412} and derived in its full form
in \cite{beis:0511}. In \cite{gleb:04Sm} the authors
consider Bethe ansatz for quantum strings and introduced the dressing phase as well as the
"symplectic" form of the charges appearing in the exponential\footnote{See \cite{gleb:04Sm}
for further details.}. The power of this
approach was further utilized in \cite{gleb:06:Sm,gleb:06ZF,BHL,janik,BES} leading to important
results.

Another step forward was done by Maldacena and Hofman \cite{Hofman:2006xt} who were able to map spin chain
"magnon" states to a specific class of rotating semiclassical strings 
on $R\times S^2$ \cite{Kruczenski:2004wg}. This result was soon generalized to magnon bound
states \cite{Dorey1,Dorey2,AFZ,MTT}, dual
to strings on $R\times S^3$ with two  and three non-vanishing angular
momenta. Magnon solutions and the dispersion relations for strings in beta-deformed backgrounds 
were obtained in \cite{Chu:2006ae, Bobev:2006fg}. The applicability of the considerations is restricted
to the case of large quantum numbers, i.e. quasiclassical approximations. Although we assume infinitely large
energy $E$ and momentum $J$, the difference $E-J$ is finite like in the case of the pp-wave limit.
Remarkably enough, due to the integrability on the gauge theory side
one is able to obtain the corresponding S-matrix explicitly \cite{BHL,BES}. 

Although similar in dealing with infinitely large quantum numbers, the two pictures, 
namely the pp-wave limit and the giant magnon sector, are a bit different. To see this one can use simple 
scaling arguments. While in the first case the angular momentum $J$ goes to infinity as $\sqrt{\l}\rightarrow\infty$ 
and the product $p.\l$ is kept fixed (for which $p$ has to be vanishing), in the magnon case
the momentum $p$ is kept fixed. In the first case we have point-like strings and we are at the bottom of the 
$E-J$ scale, while in the second case the string spike is essentially not a point-like object with higher
$E-J$. Following this logic it is natural to look for string solutions interpolating between these two sectors.
In a recent paper Maldacena and Swanson \cite{MS} suggested another limit, called near flat space (NFS) limit, 
which is interpolating between
the two regions above\footnote{One should note that the limit $p\lambda^{1/4}=cons$ was 
considered before in \cite{gleb:04Sm} in the context of \cite{GKP}, while the general discussion of this limit
on which our study is based appeared in \cite{MS}.}. In doing a Penrose limit of the geometry we perform a 
certain boost in the vicinity of 
a given null geodesics. The limit is taken in such a way that the product $p.\sqrt{\l}$ is fixed but the
space becomes of pp-wave type. The authors of \cite{MS} suggest a weaker boosting and the limit is
taken with $p^2\sqrt{\l}$ fixed. The resulting sigma model is much simpler than that of the $\axs$ superstring 
and more complicated than in the pp-wave limit. A remarkable feature of the reduced model is that,
as compared to the pp-wave case, it keeps much more information of the original theory. For instance,
it is integrable and the Lax connection can be obtained from the original one by taking the limit,
it possess the same supersymmetry algebra with a phase originating from the Hopf algebra structure
of the problem. The corresponding S matrix was also conjectured.
The worldsheet scattering of the reduced model and some other properties were
subsequently studied in \cite{Zarembo:2007nfs1,Zarembo:2007nfs2,pul}, see also \cite{Kluson:2007nfs}. 

The discussion above was focused on superstrings on $\axs$ background and its reductions
via certain limits and identifying the corresponding sectors of its dual $\N=4$ SYM theory. 
It is of importance, however, to extend these techniques
to other less symmetric backgrounds that have gauge theory duals. One step in this direction was given
in \cite{BT} where the authors considered the NFS limit of spaces where the $S^5$ part of the geometry is replaced
by certain Sasaki-Einstein spaces. For instance, it is known that spaces like $Y^{pq}$ and $L^{p,q,r}$
have $\N=1$ CFT duals. Remarkably, the near flat space limit of all these spaces is of the same type as in
\cite{MS}. Since the latter is integrable, one can hope that the corresponding theory may contain, 
at least, an integrable subsector.

In this paper we study the near flat limit reduction of string theory on
the Maldacena-Nunez background \cite{MN}\footnote{This geometry was originally obtained in the context of
non-abelian BPS monopoles in gauged supergravity and uplifted to ten dimesions in \cite{Cham-Volk}.}.
This background is 
intersting in several ways. First of all its dual is $\N=1$ gauge theory with interesting properties
and it is supposed to describe aspects of certain hadron physics \cite{Cobi} and other QCD issues.  
The paper is organized as follows. In the next section we briefly review the near flat space limit
of \cite{MS}. In Section 3 we review the Maldacena-Nunez background and its pp-wave limit. Using a
consistent truncation to a certain subsector we show that there exist magnon excitations of string theory in
the MN background. After that we perform the NFS limit (of bosonic part) of string theory in this background.
In the concluding section we comment on the result, its implications and possible developments.


\sect{The Near flat space limit of $\axs$}

In this short section we will review the procedure of taking the near flat space limit of
type IIB string theory on $\axs$ space suggested in \cite{MS}. As we already discussed in the 
introduction, the idea is to weaken the Penrose limit so that more structures of the original theory
get preserved. The first step is to find the null geodesics with respect to which the
pp-wave limit can be taken. We start with the standard metric of $\axs$ in global coordinates
\eq{
ds^2=R^2\lb[-\cosh^2\rho\, dt^2+d\rho^2+\sinh^2\rho\,d\tilde\Omega_3^2+
\cos^2\theta\,d\psi^2+\sin^2\theta\,d\Omega_3^2\rb]
}
where each of the unit 3-spheres, we take below $S^3$ from the spherical part, can be parameterized by
\eq{
d\Omega_3^2=d\f_1^2+\cos^2\f_1\,d\f_2+\sin^2\f_2\,d\f_3.
}
The light-like directions are characterized by covariantly constant null vectors
\eqs{
\nabla_\m v_\n=0, \quad v_\m v^\m=0.
}
The $S^5$ part of the null geodesics  is parameterized by $\theta=0$ and $\rho=0$, i.e.
in our case this is the equator of the five-sphere. Now we boost the worldsheet coordinates 
and expand about the geodesics ($\dot\psi=1$). This amounts to a redefinition of the fields as follows
\al{
& t=\sqrt{g}\sigma^++\frac{\tau}{\sqrt{g}} , \quad \psi=\sqrt{g}\sigma^++\frac{\chi}{\sqrt{g}} \notag \\
& \rho=\frac{z}{\sqrt{g}}, \quad \theta=\frac{y}{\sqrt{g}},
}
where we switched to light-cone worldsheat coordinates $\sigma^+,\sigma^-$. The parameter $g$
is related to the radius of $S^5$ (and $AdS_5$ as well) via $g=R^2/4\pi$.

The near flat space limit is, as in the Penrose limit, taking $g\rightarrow\infty$. In the Lagrangian
the leading terms are divergent, $g\cdot \p_-(\tau-\chi)$, but since they are total derivatives they can be dropped. 
The relevant (bosonic) part of the Lagrangian thus becomes
\eq{
S=4\lb[-\p_+\tau\,\p_-\tau+\p_+\chi\,\p_-\chi+\p_+\vec z\,\p_-\vec z+\p_+\vec y\,\p_-\vec y
 -\vec y^2\,\p_-\chi-\vec z^2\,\p_-\tau \rb]
}
where we skipped the fermionic part and used the notation $\vec z$ for coordinates in $AdS_5$ and
$\vec y$ for those in the $S^5$ part.
The reduced model has two conserved chiral currents
\al{
& j^\chi_+=\p_+\chi-\frac{\vec y^2}{2} + \text{fermions}, \quad \p_-\chi^\chi_+=0 \notag \\
&j^\tau_+=\p_+\tau-\frac{\vec z^2}{2}+ \text{fermions}, \quad \p_-\chi^\tau_+=0. 
}
The right-moving conformal invariance in preserved since its generator
\eq{
T_{--}=-(\p_\tau)^2+(\p_\chi)^2+(\p_-\vec z)^2+(\p_-\vec y)^2+\text{fermions}
}
is conserved, i.e. $\p_+T_{--}=0$. The left-moving conformal invariance however is broken, since 
$T_{++} \varpropto (j^\chi_+-j^\tau_+)$. One can still impose the Virasoro constraints
requiring
\al{
& j^\chi_++j^\tau_+=\p_+(\tau+\chi)+\frac{z^2-y^2}{2}=0, \notag \\
& T_{--}=0.
}
This can also be considered as a gauge fixing condition.

It is useful now to make a change of variables
\eq{
x^+=\sigma^+, \quad x^-=2(\tau+\chi).
}
For completeness we write down the complete gauge fixed Lagrangian
\al{
\L=& 4\Big\{\p_+\vec z\,\p_-\vec z+\p_+\vec y\,\p_-\vec y -\frac{1}{4}(\vec z^2+\vec y^2)
+(\vec y^2-\vec x^2)[(\p_-\vec z)^2+(\p_-\vec y)^2] \notag \\
& +i\psi_+\p_-\psi_+ + i\psi_-\p_-\psi_- + i\psi_-\Pi\psi_+ + i(\vec y^2-\vec x^2)\psi_-\p_-\psi_- \notag \\
& -\psi_-(\p_-z^j\Gamma^j+\p_-y^{j'}\Gamma^{j'})(z^i\Gamma^i-y^{i'}\Gamma^{i'})\psi_- \notag \\
& +\frac{1}{24}\big[\psi_-\Gamma^{ij}\psi_-\psi_-\Gamma^{ij}\psi_-
\psi_-\Gamma^{i'j'}\psi_-\psi_-\Gamma^{i'j'}\psi_\big]\Big\}.
}
In the last expression a simple rescaling of $\psi_\pm$ was used. The indices $i$ and $i'$ correspond
to the transverse directions in the anti-de Sitter and the spherical parts, respectively.


\sect{Near flat space limit of string theory in the Maldacena-Nunez background}

\subsubsection*{The background}

The background we will consider can be produced by a stack
of N D5-branes located at the origin of the transversal coordinate $\rho$ and partially
wrapping a supersymmetric cycle inside a Calabi-Yau three-fold, i.e the brane is wrapped on a 
two sphere of radius $R_2 = Ne^{2g}$. We want to preserve some supersymmetry, i.e. one must twist
the normal bundle, which is achieved by embedding the spin connection
into the R-symmetry group $SO(4)$ of D5-branes. The embedding
$U(1)_R\subset SU(2)_R\subset SO(4)$ preserves four supercharges and
thus the unwrapped sector of
the world-volume contains the $\N=1$ SYM field theory in four dimensions.
It is also assumed that in this procedure some world-volume fields become sufficiently massive 
to decouple. The resulting geometry can be summarized as follows. The metric is given by
\begin{align}
 ds^2_{str} & =e^{-\f}\lb\{dx^2_4+\a'g_sN\lb[d\rho^2+e^{2g(\rho)}\lb(d\theta_1^2+\sin^2\theta_1\,d\f_1^2\rb)
+\frac{1}{4}\sum\limits_{a=1}^3\lb(\omega^a-A^a\rb)^2\rb]\rb\} \notag\\
H^{RR} &=N\lb[-\frac{1}{4}(\omega^1-A^1)\wedge(\omega^2-A^2)\wedge(\omega^3-A^3)+\frac{1}{4}
\sum\limits_{a=1}^3F^a\wedge (\omega^a-A^a)\rb], 
\label{background}
\end{align}
where
\begin{align}
& A=\frac{1}{2}[\sigma^1\,a(\rho)\,d\theta_1+\sigma^2\,a(\rho)\sin\theta_1\,d\phi_1+
\sigma^1\,\cos\theta_1\,d\phi_1], \label{ungaugedA}\\
& e^{2\f}=e^{2\f_0}\frac{2e^{g(\rho)}}{\sinh(2\rho)}, \quad
e^{2g(\rho)}=\rho\,\coth(2\rho)-\frac{\rho^2}{\sinh^2(2\rho)}-\frac{1}{4}
\quad a(\rho)=\frac{2\rho}{\sinh(2\rho)}.
\end{align}
In the above expression we introduced the notations for the left-invariant $SU(2)$ one-forms
\eq{
 \frac{i}{2}\omega^a\sigma^a=dg\,g^{-1}, 
\quad g=e^{\frac{i}{2}\psi\sigma^3}e^{\frac{i}{2}\theta_2\sigma^1}e^{\frac{i}{2}\phi_2\sigma^3}, \notag \\
}
or, explicitly
\begin{align}
 & \omega^1+i\omega^2=e^{-i\psi}(d\theta_2+i\sin\theta_2\,d\f_2), \quad
\omega^3=d\psi+\cos\theta_2\,d\f_2; \notag \\
& \omega^1\omega^1 +\omega^2\omega^2 + \omega^3\omega^3=d\theta_2^2+d\f_2^2+d\psi^2
+2\cos\theta_2 d\f_2d\psi \label{sum-omega2}
\end{align}
For later purposes it will be useful to study the behavior of the geometry near the origin. 
To do that let us expand the factor $a(\rho)$ for small $\rho$ (where the warp factor, which can be
considered as a potential, has its minimum)
\eq{
a(\rho)=\frac{2\rho}{\sinh(2\rho)}\approx 1-\frac{2}{3}\rho^2+\O(\rho^4). \label{exp-a}
}
Substituting in \eqref{ungaugedA} we find
\eq{
A=-idh\,h^{-1}+\O(\rho^2), \label{pure-g}
}
where
\eqs{
h=e^{i\sigma^1\theta_1/2}e^{i\sigma^3\phi_1/2}, \quad \text{($\sigma^i$ are the Pauli matrices).}
}
The fact that at $\rho=0$ the gauge field is pure gauge was observed already in the original paper by Maldacena and Nunez
\cite{MN}.
Therefore one can simplify a bit the considerations expanding the background for small $\rho$ and performing 
the following gauge transformation
\eq{
A\longrightarrow h^{-1}Ah+dh\,h^{-1}=\lb(-\frac{1}{3}\rho^2\rb)h^{-1}
\lb[\sigma^1\,d\theta_1+\sigma^2\,\sin\theta_1 d\phi_1\rb]h \label{gaugedA-0}
}
where we used \eqref{pure-g} and \eqref{exp-a}.
Since the Pauli matrices ($\sigma^1,\sigma^2$) transform as
\als{
& h^{-1}\sigma^1\,h=\sigma^1\cos\phi_1+\sigma^2\sin\phi_1 \\
& h^{-1}\sigma^2\,h=\sigma^1(-\sin\phi_1\cos\theta_1)+\sigma^2(\cos\theta_1\cos\phi_1)+
\sigma^3(\sin^2\theta_1)
}
we find from \eqref{gaugedA-0} the following expression for the gauge field (up to $\rho^4$)
\eq{
 A= \lb(-\frac{1}{3}\rho^2\rb)\hat A+\O(\rho^4), \label{expres-exp}
}
where  $\hat A$ is given by
\ml{
\hat A=\sigma^1(\cos\phi_1d\theta_1-\cos\theta_1\sin\theta_1\sin\f_1d\f_1) \\
 +\sigma^2(\sin\f_1d\theta_1+\cos\theta_1\sin\theta_1\cos\f_1 d\f_1)
+\sigma^3(\sin^2\theta_1 d\f_1). \notag
}
With this simplification the last term of the metric becomes
\eq{
\frac{1}{4}\sum\limits_{a=1}^3(\omega^a-A^a)^2=\frac{1}{4}\sum\limits_{a=1}^3
\lb[(\omega^a)^2+\frac{2}{3}\rho^2\omega^a\hat A^a\rb]+\O(\rho^4) \label{expres-fin}
}

\subsubsection*{The pp-wave limit in brief}

Here we will review the  pp-wave limit following closely the approach given in \cite{Cobi}. To obtain the pp-wave limit of
this geometry one has to find null geodesics. The geometry is complicated, it is actually a fibration of $S^3$ 
over $S^2$ with radius $e^{2g(\rho)}$. To simplify the considerations it will be convenient to take a look 
first at the behavior of the radial 
part. One can easily check that $\rho=0$ is a solution to the equations of motion (EOM). Some more 
observations are important: \\
a) In conformal gauge $t=\kappa\tau$ and then the warp factor $e^{-\phi(\rho)}$ serves as a potential for the dynamics of
$\rho$. One can easily check that at $\rho=0$ this potential has a minimum and therefore this solution
is stable. \\
b) Setting $\rho=0$  means that the radius $e^{2g(\rho)}$ of the internal $S^2$ sphere (parameterized
by $d\theta_1^2+\sin^2\theta_1\,d\phi_1^2$) shrinks to zero and this sector decouples. \\
c) The only contribution from the compact part of the geometry is \eqref{sum-omega2}
\eqs{
\sum\limits_{a=1}^3 \lb(\omega^a\rb)^2=d\theta_2^2+d\f_2^2+d\psi^2+2\cos\theta_2 d\f_2d\psi.
}

Now it is easy to find the null geodesics we are looking for. It is actually the equator of the internal
$S^3$ parameterized for instance by $\theta_2=0$. Then we find
\eq{
dt^2=d\vf^2, \quad \text{where} \,\,\, \vf=\frac{1}{2}(\phi_2+\psi).
}
The latter equality means that we have two global quantum numbers, the energy $E$ and the angular 
momentum, say $J$.\\
In order to obtain the pp-wave limit we have to zoom all the curvature in the vicinity of
this null geodesics \cite{Cobi}. The resulting pp-wave geometry is
\ml{
ds^2=-2dx^+dx^--m_0^2\lb(\frac{1}{9}u_1^2+\frac{1}{9}u_2^2+v^2\rb)(dx^+)^2+
d\vec x^2\\
+dz^2+du_1^2+du_2^2+dv_1^2+dv_2^2,
}
where the following scalings and notations are used
\als{
& m_0=\frac{1}{\a'g_sN}, \quad \rho=\frac{r}{\sqrt{g}},\quad \theta_2=\frac{2m_0}{\sqrt{g}}v,
\quad x^i\rightarrow \frac{x^i}{\sqrt{g}}, \\
& \vf_1=\f_1+\frac{1}{3}\vf, \quad \vf_2=\f_2-\vf, \\
& dr^2+r^2(d\theta_2^2+\sin^2\theta_2 d\vf_1^2)\,\,\rightarrow \,\,du_1^2+du_2^2+dz^2, \\
& dv^2+v^2d\vf_2^2\,\, \rightarrow\,\, dv_1^2+dv_2^2.
}
The RR field strength in the Penrose limit scales as
\eqs{
H^{RR}=g\,\tilde H^{RR}
}
so, since $\omega^{1,2}\sim 1/\sqrt{g}$ and $\omega^2\sim \O(g^0)$, only the terms 
$\omega^1 \wedge\omega^2 \wedge \omega^3$ and $F^3\wedge \omega^3$ will contribute
in the limit $g\rightarrow \infty$ (others are suppressed by $1/\sqrt{g}$),
\eq{
H^{RR}=2-m_0\,dx^+\wedge[dv_1\wedge dv_2+\frac{1}{3}du_1\wedge du_2].
}
The pp-wave reduced model is exactly solvable, but in the process of taking the
Penrose limit some of the key features of the original model get lost.

\subsubsection*{A subsector with giant magnon solutions}

Another simplification of the theory comes with the idea to look for
classical solutions with large quantum numbers and to study the fluctuations about them. 
As we discussed in the Introduction, such an important  class of solutions is the class of so 
called ``giant magnon'' solutions \cite{Hofman:2006xt}. 
Here we are going to show their existence at least in one subsector of our theory.

We start again with the solution $\rho=0=\vec x$. As we discussed above, in this case
the Lagrangian greatly simplifies and reduces to a sigma model on $R_t\times S^3$
\ml{
\L=-\frac{\sqrt{\l}}{4\pi}\Big\{-\p_\a t\p^\a t + \frac{m_0^2}{4}\big[\p_\a \theta_2\p^\a \theta_2 
+\p_\a \f_2\p^\a \f_2+\p_\a \psi\p^\a \psi+2\cos\theta_2\,\p_\a\f_2
\p^\a\psi\big]\Big\}
}
where in conformal gauge we are using the Weyl invariant metric $\gamma^{\a\b}=\sqrt{h}h^{\a\b}=diag(-1,1)$.
Defining
\eq{
\vf=\frac{1}{2}(\f_2+\psi), \quad \e=\frac{1}{2}(\f_2-\psi),\quad \theta_2=\theta
}
we obtain the Lagrangian in the form
\ml{
\L=\,\,(\p_\tau t)^2-(\p_\sigma t)^2+(\p_\sigma\theta)^2-(\p_\tau\theta)^2 \\
 + \sin^2\theta\big((\p_\sigma\vf)^2-(\p_\tau\vf)^2\big)+
\cos^2\theta\big((\p_\sigma\e)^2-(\p_\tau\e)^2\big).
}
The reduced theory is supplied by the Virasoro constraint
\eq{
\p_\sigma\theta^2+\p_\tau\theta^2+\sin^2\theta(\p_\sigma\vf^2+\p_\tau\vf^2)
+\cos^2\theta(\p_\sigma\e^2+\p_\tau\e^2)=1
}
where we used the residual symmetry to set $t=\tau$.
The following ansatz leads to giant magnon solutions
\al{
& \theta=\theta(y), \quad y=c\sigma-d\,\tau, \quad c^2-d^2=1, \\
& \vf=\tau+g_1(y), \quad \e=\n\tau+g_2(y).
\label{mag-ansatz}
}
In this parameterization the Virasoro constraints become
\ml{
(d^2+c^2){\theta'}^2+\sin^2\theta\big[c^2{g_1'}^2+(1-d\,{g_1'}^2)\big]
+\cos^2\theta\big[c^2{g_2'}^2+(\n-d\,{g_1'}^2)\big]=1
}
(the prime stands for a derivative with respect to $y$) and the equations of motion (EOM) read
\al{
& g_1'=d\cot^2\theta   \\
& g_2'=-\n d.
}
Combining the Virasoro constraints with the EOM we end up with
\eq{
\theta'=\cos\theta\sqrt{1-\n^2 c^2-d^2\cot^2\theta}.
}
This equation can be easily solved and the solutions is
\eq{
\cos\theta=\sqrt{\frac{1-\n^2 c^2}{c^2-\n^2 c^2}}\,\frac{1}{\cosh\sqrt{1-\n^2 c^2 y}}.
}
The ansatz \eqref{mag-ansatz} ensures the existence of three global charges
\al{
& E=\frac{\l}{2\pi}\int\limits_{-\infty}^\infty d\sigma, \\
& J_1=\frac{\l}{2\pi}\int\limits_{-\infty}^\infty d\sigma\sin^2\theta\,\big(1-g_1'(y)\big), \\
& J_2=\frac{\l}{2\pi}\int\limits_{-\infty}^\infty d\sigma\cos^2\theta\,\big(\n-g_2'(y)\big).
}
A little algebra leads to the dispersion relations for the magnon bound state found in \cite{Dorey1} and rederived
for the string sigma model on $R_t\times S^3$ in \cite{AFZ, MTT},
\eq{
E-J_1=\sqrt{J_2^2+\frac{\l}{\pi^2}\sin\frac{p}{2}}.
}
Other more general approaches are to reduce the dynamical system to Neumann-Rosochatius integrable system,
as is given in \cite{Kruczenski:2006pk, DR-spikes}, or to use Pohlmeyer reduction of the $O(4)$ sigma model
\cite{Grigoriev:2007pohl}. 
The latter ends up with complex the
Sine-Gordon model which, using so called helical coordinates \cite{cSG}, determines the required type of solutions 
through the Lam\'{e} equation, while the first approach enables the represention of the solutions through theta functions.

\subsubsection*{Near flat space limit}

Let us start with the following redefinition of the coordinates 
\eq{
\vf=\frac{\psi+\phi_2}{2} \,\, \Longrightarrow \,\, \psi=2\vf-\phi_2
}
and the rescaling ($m_0=1/\sqrt{\a'g_sN}$ is kept fixed)
\eq{
\rho=\frac{m_0}{\sqrt{g}}r, \quad \theta_2=\frac{2m_0}{\sqrt{g}}v. \label{rescal-1}
}
The components $\omega^a$ are explicitly read off from \eqref{sum-omega2}
\al{
& \omega^1=\cos\psi d\theta_2+\sin\psi\sin\theta_2 d\phi_2 \label{omega-1} \\
& \omega^2=-\sin\psi d\theta_2+\cos\psi\sin\theta_2 d\phi_2 \label{omega-2} \\
& \omega^3= d\psi+\cos\theta_2 d\phi_2 \label{omega-3}
}
or, after using the ansatz above and taking a large $g$ expansion,
\al{
& \omega^1=\frac{2m_0}{\sqrt{g}}\lb[\cos(2\vf-\phi_2)dv+v\sin(2\vf-\phi_2)d\phi_2\rb] 
+\O(g^{-3/2}) \label{omega-1a}, \\
& \omega^2=\frac{2m_0}{\sqrt{g}}\lb[-\sin(2\vf-\phi_2)dv+v\cos(2\vf-\phi_2)d\phi_2\rb]
+\O(g^{-3/2})  \label{omega-2a}, \\
& \omega^3=2d\vf-2\frac{m_0^2v^2}{g} d\phi_2 + \O(g^{-2}) \label{omega-3a}.
}
Then, after the rescaling \eqref{rescal-1}, from \eqref{sum-omega2} we find for the part of the metric 
involving $\omega^a$ the expression
\eq{
\frac{1}{4m_0^2}\sum\limits_{a=1}^3(\omega^a)^2=\frac{1}{m_0^2}d\vf^2+\frac{1}{g}(dv^2+v^2 d\phi_2^2)
-\frac{2v^2}{g}d\vf d\phi_2 + \O(g^{-2}).
\label{omega2-2}
}
Now according to \eqref{expres-exp} and \eqref{expres-fin} we have to compute $\omega^a\hat A^a$.
For $\omega^1\hat A^1+\omega^2\hat A^2$ we find
\ml{
\sum\limits_{i=1}^2\omega^i\hat A^i= \frac{2m_0}{\sqrt{g}}\Big[
\cos\eta\, dvd\theta_1-\sin\eta\sin\theta_1\cos\theta_1\, dv d\f_1 \\
+v\sin\eta\, d\f_2 d\theta_1+v\cos\eta\sin\theta_1\cos\theta_1\, d\f_2 d\f_1 \Big],
\label{F-1}
} 
where
\eq{
\e=2\vf+\f_1-\f_2. \label{c-phase}
}
{}From the remaining term $\omega^3\hat A^3$ we get
\eq{
\omega^3\hat A^3=2\sin^2\theta_1\,d\vf d\f_1-\frac{2m_0^2 v^2}{g}\sin^2\theta_1\,d\f_2 d\f_1
}
Substituting the above expressions into the metric we find the following final expression 
\ml{
ds^2=e^{-\f}\Big\{-dt^2+\frac{1}{m_0^2}d\vf^2+\frac{1}{g}
\Big[d\vec x^2+dr^2+r^2(d\theta_1^2+\sin^2\theta_1\,d\f_1^2) \\
+(dv^2+v^2\,d\f_2^2)-2v^2\,d\f_2d\vf+\frac{2r^2}{3}\sin^2\theta_1\,d\f_1d\vf\Big] +F \Big\},
\label{metric-2}
}
where $F$ is
\ml{
F=\frac{m_0r^2}{3g\sqrt{g}}\Big[
\cos\e\, dvd\theta_1-\sin\e\sin\theta_1\cos\theta_1\, dv d\f_1 \\
+v\sin\e\, d\f_2 d\theta_1+v\cos\e\sin\theta_1\cos\theta_1\, d\f_2 d\f_1 \Big]
\equiv \frac{m_0r^2}{3g\sqrt{g}} \tilde F.
\label{F-2}
}
It will be convenient for what follows to diagonalize the part of the metric,
which is given in \eqref{metric-2}, with respect to the variables $\vf,\f_1,\f_2$. This
can be done by redefining the angles $\f_1$and $\f_2$ as follows
\eq{
\f_1=\vf_1-\frac{1}{3}\vf, \quad \f_2=\vf_2+\vf. \label{redef-f}
}
Substituting back in \eqref{metric-2} we find
\ml{
ds^2=e^{-\f}\Big\{-dt^2+d\vf^2+\frac{1}{g}
\Big[d\vec x^2+dr^2+r^2(d\theta_1^2+\sin^2\theta_1\,d\vf_1^2) \\
+(dv^2+v^2\,d\vf_2^2)-m_0^2(v^2+\frac{r^2}{9}\sin^2\theta_1)\,d\vf^2 \Big] +F \Big\},
\label{metric-3}
}
where we also rescaled $\phi\rightarrow m_0\phi$.


Before we proceed further with the near flat space limit, let us analyze the behavior of the
prefactor $e^{-\phi}$  multiplying the whole metric. We are interested in the small $\rho$ 
expansion of
\eqs{
e^{2\phi}=e^{2\phi_0}\frac{2e^{g(\rho)}}{\sinh(2\rho)}
}
with
\eqs{
e^{2g(\rho)}=\rho\,\coth(2\rho)-\frac{\rho^2}{\sinh^2(2\rho)}-\frac{1}{4}.
}
It is a simple exercise to find the expansion of  $e^{2\phi(\rho)}$ for small 
$\rho$ which gives for the prefactor the expression
\eq{
e^{-\f}=e^{-\f_0}\lb(1+\frac{4}{9}\rho^2 +\O(\rho^4)\rb). \label{prefactor}
}

The conclusion is that, multiplied by $g$,  the two terms in the small $\rho$ expansion
\eqref{prefactor} will contribute  ($\rho^2\sim 1/g$) to the NFS limit.

Now we return to the analysis of the near flat space limit. {}From \eqref{metric-3} and the
expansion \eqref{prefactor} it follows that one must ensure finiteness of \eqref{metric-3}
when it is multiplied by the overall factor of $g$. The finiteness condition forces the following ansatz
for the near flat space limit
\al{
&  x^i\equiv x^i, \quad \theta_1\equiv \theta_1,\quad  \vf_1 \equiv \vf_1,
\quad \vf_2 \equiv \vf_2  \label{1.1}\\
& \theta_2 =\frac{2m_0}{\sqrt{g}}v, \quad  \rho=\frac{m_0}{\sqrt{g}}r, \label{1.2}\\
& t= k_t\sqrt{g}\sigma^++\frac{\tau}{\sqrt{g}},\quad
\vf  =k\sqrt{g}\sigma^++\frac{\chi}{\sqrt{g}} \label{1.3}
}
It is easy to check that the explicit part of \eqref{metric-3} is finite since
\eqs{
dt^2=\lb(k_t\,\p_-\tau + \frac{\p_+\tau\p_-\tau}{g}\rb)d\sigma^+d\sigma^-, \quad
d\vf^2= \lb(k\,\p_-\chi + \frac{\p_+\chi\p_-\chi}{g}\rb)d\sigma^+d\sigma^-
}
and the first terms are total derivatives - they can be integrated out.

One should note that in contrast to the non-warped products of geometries like
$AdS_5\times S^5$ \cite{MS} or Sasaki-Einstein metrics $T^{p,q}$, $Y^{p,q}$ \cite{BT}, 
one also gets a contribution from the warp factor, it arises from the second term of 
\eqref{prefactor}, and it is
\eq{
\frac{4}{9}\,r^2(-k_t\,\p_-\tau+k\,\p_-\chi). \label{add-contr}
}
Let us analyze the $F$-term in \eqref{metric-3} whose explicit form is given
in \eqref{F-2}. This term is proportional to $g^{-3/2}$ but after the redefinitions 
\eqref{redef-f} and overall multiplication of the
Lagrangian by $g$ terms proportional to $\sqrt{g}$ appear. They come from
the terms proportional to
\eqs{
dvd\vf, \quad d\theta_1 d\vf, \quad (d\vf_1-\frac{1}{3}d\vf_2)d\vf.
}
Here $v,\vf_1$ and $\vf_2$ are finite but $\vf$ is proportional to $\sqrt{g}$
and so there may be finite contributions. However
all the terms are multiplied by $\cos\e$ or $\sin\e$. But due to \eqref{c-phase}
$\e=2k/3\sqrt{g}\sigma^++(\text{finite part})$ and therefore it can be decomposed to well
defined finite terms multiplied by $\cos(2k/3\sqrt{g}\sigma^+)$ or $\sin(2k/3\sqrt{g}\sigma^+)$.
In the limit $g\rightarrow\infty$ one might expect that these fast oscillating factors vanish
because their average values are zero. Actually this can be explicitly demonstrated. We have terms of 
the form
\eq{
\mathbf{D}:=f(v,\theta_1,\vf_1,\vf_2).k\sqrt{g}\lb\{
\begin{matrix} \cos(\frac{2}{3}k\sqrt{g}\sigma^+) \\ \sin(\frac{2}{3}k\sqrt{g}\sigma^+) 
\end{matrix}\rb\}
\label{oscill-part}
}
where $f(v,\theta_1,\vf_1,\vf_2)$ is a well define finite function of its arguments.
Instead of \eqref{oscill-part} one can consider the more general expression
\eq{
\mathbf{D}_s:=f(x^i,\theta_j,z)\,g^s\,\lb\{
\begin{matrix} \cos(\sqrt{g}\sigma^+) \\ \sin(\sqrt{g}\sigma^+)
 \end{matrix}\rb\},
 \quad s\in \frac{1}{2}{\mathbb Z} \label{oscill-part-a}
}
where $f(x^i,\theta_j,z)$ is independent of $g$. This can be written also as
\al{
\mathbf{D}_s & =f(x^i,\theta_j,z)g^{s-\frac{1}{2}}\p_+\lb\{
\begin{matrix} \sin(\sqrt{g}\sigma^+) \\ 
-\cos(\sqrt{g}\sigma^+) 
\end{matrix}\rb\} \notag \\
 & =
\p_+\lb\{f(x^i,\theta_j,z)\,g^{s-\frac{1}{2}}\,
\begin{matrix} \cos(\sqrt{g}\sigma^+) \\ 
\sin(\sqrt{g}\sigma^+) \end{matrix}\rb\}-
\p_+f(x^i,\theta_j,z)\,g^{s-\frac{1}{2}}\,
\lb\{\begin{matrix} \cos(\sqrt{g}\sigma^+) \\ 
\sin(\sqrt{g}\sigma^+) \end{matrix}\rb\}  \notag \\
& =
\p_+\lb\{f(x^i,\theta_j,z)\,g^{s-\frac{1}{2}}\,
\begin{matrix} \cos(\sqrt{g}\sigma^+) \\ 
\sin(\sqrt{g}\sigma^+) \end{matrix}\rb\}-\mathbf{D}_{s-\frac{1}{2}} .
\label{oscill-part-b}
}
The first term in the last equality is a total derivative and can be integrated out. The second
term  is now of order $s-\frac{1}{2} $. Repeating this procedure $2s$ or more times we will end up with
$\mathbf{D}_\a$ where $\a< 0$ and therefore the $F$-term vanishes in the limit 
$g\rightarrow\infty$ (up to total derivatives).

In our case $\mathbf{D}$ is just $\mathbf{D}_{1/2}$, so this terms does not contribute to
the near flat space limit. Therefore, since the whole $F$-term vanishes,
the contributions to the Lagrangian come only from the explicit part of the metric \eqref{metric-3}.

\subsubsection*{The Lagrangian}

Now, with the large $g$ analysis of the metric at hand, we are in a position to write down the
Lagrangian of the reduced model. It takes the form
\ml{
\L=2e^{-\phi_0}\Big\{-\p_-\tau\p_+\tau-k_t \frac{4 m_0^2}{9}\,z^2\,\p_-\tau +
k\frac{4 m_0^2}{9}\,z^2\,\p_-\chi+
\p_-\chi\p_+\chi+ \\
+\p_-x^i\p_+x^i+\p_-r\p_+r +r^2\p_-\theta_1\p_+\theta_1+r^2\sin^2\theta_1\,\p_-\vf_1\p_+\vf_1 \\
+\p_-v\p_+v+v^2\,\p_-\vf_2\p_+\vf_2-km_0^2(r^2+\frac{z^2}{9}\sin^2\theta_1)\,\p_-\chi
\Big\}
}
The Virasoro constraints fix $k=k_t$, which we set to one. With some obvious coordinate redefinition we obtain
\begin{align}
\Lag= & 2e^{-\phi_{0}}\bigg(-\frac{4 m_0^2}{9}z^{2}\partial_{-}\tau-\partial_{-}\tau\partial_{+}\tau+\frac{4 m_0^2}{9}z^{2}\partial_{-}\chi+\partial_{-}\chi\partial_{+}\chi+\partial_{+}\vec{y}
\partial_{-}\vec{y}+\partial_{+}\vec{z}\partial_{-}\vec{z}+\nonumber\\
 & +\partial_{+}\vec{r}\partial_{-}\vec{r}-m_0^2\left(r^{2}+
\frac{1}{9}\left(z_{2}^{2}+z_{3}^{2}\right)\right)\partial_{-}\chi\bigg)
+\Ord\left(g^{-\frac{1}{2}}+\partial\right).\label{lagrangian}
\end{align}

The equations of motion following from the reduced Lagrangian
\eqref{lagrangian} are
\begin{align*}
0= & \partial_{+}\partial_{-}\tau+\frac{2 m_0^2}{9}\partial_{-}z^{2}\\
0= & \partial_{+}\partial_{-}\chi+\frac{2 m_0^2}{9}\partial_{-}z^{2}
-\frac{m_0^2}{2}\partial_{-}\left(r^{2}+\frac{1}{9}\left(z_{2}^{2}+z_{3}^{2}\right)\right)\\
0= & \partial_{+}\partial_{-}\vec{y}\\
0= & \partial_{+}\partial_{-}z_{i}+\frac{4 m_0^2}{9}\, z_{i}(\partial_{-}\tau-\partial_{-}\chi)+\frac{m_0^2}{9}
\left(\delta_{i,2}z_{2}+\delta_{i,3}z_{3}\right)\partial_{-}\chi\qquad i=1,2,3\\
0= & \partial_{+}\partial_{-}r_{i}+m_0^2\,r_{i}\,\partial_{-}\chi\qquad i=1,2
\end{align*}
As in \cite{MS}, from the \EOM~we obtain the following chiral conserved currents
\begin{align*}
j_{+}^{\tau}= & \partial_{+}\tau+\frac{2 m_0^2}{9}z^{2},\\
j_{+}^{\chi}= & \partial_{+}\chi+\frac{2 m_0^2}{9}z^{2}-\frac{m_0^2}{2}\left(r^{2}
+\frac{1}{9}\left(z_{2}^{2}+z_{3}^{2}\right)\right).
\end{align*}
The next step is to analyze the compatibility of the NFS limit with the Virasoro
constraints and the symmetries of the reduced model.

\subsubsection*{Virasoro constraints}

Our starting point is the expression for the energy-momentum tensor in the original
theory
\begin{align*}
T_{++}= & G_{\mu\nu}\partial_{+}X^{\mu}\partial_{+}X^{\nu},\qquad T_{--}
=G_{\mu\nu}\partial_{-}X^{\mu}\partial_{-}X^{\nu}.
\end{align*}
The next step is to use the rescaling (\ref{1.1}-\ref{1.3}) and the redefinitions used in the previous section
to obtain the corresponding expression for the reduced model. The result is
\begin{align*}
T_{++}\rightarrow & e^{-\phi_{0}}\bigg(-\partial_{+}\tau+\partial_{+}\chi-
\frac{m_0^2}{2}\left(r^{2}+
\frac{1}{9}\left(z_{2}^{2}+z_{3}^{2}\right)\right)\bigg),\\
T_{--}\rightarrow & \frac{1}{g}e^{-\phi_{0}}\bigg(-\partial_{-}\tau\partial_{-}\tau+
\partial_{-}\chi\partial_{-}\chi+\partial_{-}y_{i}\partial_{-}y_{i}+
\partial_{-}\vec{z}\partial_{-}\vec{z}+\partial_{-}\vec{r}\partial_{-}\vec{r}\bigg).
\end{align*}
{}From the expression for $T_{++}$ ($\varpropto j_+^\chi-j_+^\tau$) we see that the same situation as in \cite{MS}
is realized.
The left-moving conformal symmetry is broken and gets replaced by two chiral symmetries which are 
generated by $j_+^\tau$ and $j_+^\chi$. The right-moving conformal symmetry, corresponding to
parameterization $\sigma^-\rightarrow f(\sigma^-)$ with arbitrary $f$, remains unbroken. Its generator
is
\begin{align*}
 T_{--}:=&-\partial_{-}\tau\partial_{-}\tau+
\partial_{-}\chi\partial_{-}\chi+\partial_{-}y_{i}
\partial_{-}y_{i}+\partial_{-}\vec{z}\partial_{-}\vec{z}+\partial_{-}\vec{r}\partial_{-}\vec{r}.
\end{align*}

\subsubsection*{Gauge fixed action}

To gauge fix the action we use the same procedure as applied in \cite{MS}. 
We impose the following conditions
\begin{align*}
T_{--}= & 0,\\
j_{+}^{\tau}+j_{+}^{\chi}= & 0=\partial_{+}\tau+\partial_{+}\chi+\frac{4 m_0^2}{9}z^{2}-\frac{m_0^2}{2}
\left(r^{2}+\frac{1}{9}\left(z_{2}^{2}+z_{3}^{2}\right)\right),
\end{align*}
and choose the coordinates
\begin{align*}
x^{+}\equiv & \sigma^{+},\qquad x^{-}\equiv2\left(\tau+\chi\right),\end{align*}
so that the derivatives become
\begin{align*}
\partial_{\sigma^{+}}= & \partial_{x^{+}}+\left[-\frac{8 m_0^2}{9}z^{2}+m_0^2\left(r^{2}+
\frac{1}{9}\left(z_{2}^{2}+z_{3}^{2}\right)\right)\right]\partial_{x^{-}},
\qquad\partial_{\sigma^{-}}=2\left[\partial_{\sigma^{-}}\left(\tau+\chi\right)\right]\partial_{x^{-}}.
\end{align*}

With this redefinitions we can rewrite the \EOM~for $y$, $z$ and $r$.
\begin{align*}
0= & \partial_{x^{-}}\left(\partial_{x^{+}}+\left[-\frac{8 m_0^2}{9}z^{2}+m_0^2\left(r^{2}+
\frac{1}{9}\left(z_{2}^{2}+z_{3}^{2}\right)\right)\right]\partial_{x^{-}}\right)\vec{y},\\
0= & \partial_{x^{-}}\left(\partial_{x^{+}}+\left[-\frac{8 m_0^2}{9}z^{2}+m_0^2
\left(r^{2}+\frac{1}{9}\left(z_{2}^{2}+z_{3}^{2}\right)\right)\right]\partial_{x^{-}}\right)z_{i}+\\
&+z_{i}\frac{8 m_0^2}{9}\left(\partial_{x^{-}}\vec{y}\partial_{x^{-}}\vec{y}+
\partial_{x^{-}}\vec{z}\partial_{x^{-}}\vec{z}+\partial_{x^{-}}\vec{r}\partial_{x^{-}}\vec{r}\right)+\\
 &+\frac{m_0^2}{9}\left(\delta_{i,2}z_{2}+\delta_{i,3}z_{3}\right)\left(\frac{1}{4}-
\partial_{x^{-}}\vec{y}\partial_{x^{-}}\vec{y}-\partial_{x^{-}}\vec{z}\partial_{x^{-}}\vec{z}
-\partial_{x^{-}}\vec{r}\partial_{x^{-}}\vec{r}\right),\qquad i=1,2,3 \notag \\
\end{align*}
\begin{align*}
0= & \partial_{x^{-}}\left(\partial_{x^{+}}+\left[-\frac{8 m_0^2}{9}z^{2}+m_0^2
\left(r^{2}+\frac{1}{9}\left(z_{2}^{2}+z_{3}^{2}\right)\right)\right]\partial_{x^{-}}\right)
r_{i} \notag \\
&+m_0^2r_{i}\left(\frac{1}{4}-\partial_{x^{-}}\vec{y}\partial_{x^{-}}\vec{y}
-\partial_{x^{-}}\vec{z}\partial_{x^{-}}\vec{z}-\partial_{x^{-}}\vec{r}\partial_{x^{-}}\vec{r}\right),
\notag \\
 & \qquad i=1,2 
\end{align*}
where we used
\eq{
T_{--}=  0\quad
\Rightarrow \quad \frac{1}{2}\partial_{x^{-}}\left(\tau-\chi\right)= 
\partial_{x^{-}}y_{i}\partial_{x^{-}}y_{i}+\partial_{x^{-}}\vec{z}
\partial_{x^{-}}\vec{z}+\partial_{x^{-}}\vec{r}\partial_{x^{-}}\vec{r}.
}
These equations can be though of as \EOM's derived from the following effective Lagrangian
\begin{align}
L_{\txt{eff}}= & \partial_{+}\vec{y}\partial_{-}\vec{y}+\left[-\frac{8 m_0^2}{9}z^{2}+m_0^2\left(r^{2}+
\frac{1}{9}\left(z_{2}^{2}+z_{3}^{2}\right)\right)\right]\partial_{-}\vec{y}\partial_{-}\vec{y} \notag \\
 & +\partial_{+}\vec{z}\partial_{-}\vec{z}+\left[-\frac{8 m_0^2}{9}z^{2}+m_0^2
\left(r^{2}+\frac{1}{9}\left(z_{2}^{2}+z_{3}^{2}\right)\right)\right]\partial_{-}\vec{z}\partial_{-}\vec{z} \notag \\
 & +\partial_{+}\vec{r}\partial_{-}\vec{r}+\left[-\frac{8 m_0^2}{9}z^{2}+m_0^2
\left(r^{2}+\frac{1}{9}\left(z_{2}^{2}+z_{3}^{2}\right)\right)\right]\partial_{-}\vec{r}\partial_{-}\vec{r}\notag\\
 & -\frac{m_0^2}{9}\frac{z_{2}^{2}+z_{3}^{2}}{4}-m_0^2\frac{r^{2}}{4}.
\end{align}
This Lagrangian is much simpler than the Lagrangian we started with. Although simplified, it is still has
 complicated structure and hopefully incorporates some of the key properties of the original theory.

\subsubsection*{The softly broken model}

Here we will extend the consideration above to the case of the softly broken MN model 
\cite{soft-brok} whose pp-wave limit and some quasi classical properties were studied 
in \cite{cotrone-soft}. The general form of the background is the same as in 
(\ref{background}) and (\ref{ungaugedA}),
but in contrast to the supersymmetric case, the explicit form of the functions 
$\phi(\rho), g(\rho)$ and $a(\rho)$ are unknown. As we have seen
above, to study the NFS limit we only need the behavior of these functions in the limit 
$\rho \rightarrow 0$, which fortunately is known. Their small $\rho$ expansions are given by
\al{
a(\rho)&=1-b^2\rho^2+\cdots, \notag \\
e^{g(\rho)}&=\rho-(\frac{b^2}{4}+\frac{1}{9})\rho^3+\cdots,
\label{soft-asymp} \\
\phi(\rho)&=\phi_0+(\frac{b^2}{4}+\frac{1}{3})\rho^2+\cdots. \notag 
}
where the range of $b\in(0,2/3]$ is determined from the requirements for regularity
of the solution and matching suitable UV asymptotes. Note that the supersymmetric solution
corresponds to $b=2/3$\phantom{x}\footnote{For more details see \cite{soft-brok, cotrone-soft} and references
therein.}. From \eqref{soft-asymp} one can see that the procedure we developed for the supersymmetric case
can be directly applied to the softly broken case. Indeed, using that in this case
\eq{
e^{-\phi}=e^{-\phi_0}\lb(1+(\frac{b^2}{4}+\frac{1}{3})\rho^2+\O(\rho^4)\rb),
}
and repeating the above steps we obtain the following Lagrangian
\begin{multline}
\Lag=  2e^{-\phi_{0}}\bigg(-m_0^2\,z^2(\frac{b^2}{4}+\frac{1}{3})\partial_{-}\tau-
\partial_{-}\tau\partial_{+}\tau+m_0^2\,z^2(\frac{b^2}{4}+\frac{1}{3})
\partial_{-}\chi+\partial_{-}\chi\partial_{+}\chi  \\ 
 +\partial_{+}\vec{y}\partial_{-}\vec{y}+\partial_{+}\vec{z}\partial_{-}\vec{z}+
  +\partial_{+}\vec{r}\partial_{-}\vec{r}-m_0^2\left(r^{2}+
\frac{1}{9}\left(z_{2}^{2}+z_{3}^{2}\right)\right)\partial_{-}\chi\bigg)\\ 
 +\Ord\left(g^{-\frac{1}{2}}+\partial\right).\label{lagrangian-soft}
\end{multline}
Analyzing the EOM one obtains the following chiral conserved currents
\begin{align*}
j_{+}^{\tau}= & \partial_{+}\tau+\frac{m_0^2}{2}(\frac{b^2}{4}+\frac{1}{3})z^{2},\\
j_{+}^{\chi}= & \partial_{+}\chi+\frac{m_0^2}{2}(\frac{b^2}{4}+\frac{1}{3})z^{2}-\frac{m_0^2}{2}\left(r^{2}
+\frac{1}{9}\left(z_{2}^{2}+z_{3}^{2}\right)\right).
\end{align*}
We can again further gauge fix the action using
\als{
T_{--}=0, \\
j_{+}^{\tau}+j_{+}^{\chi}= & 0=\partial_{+}\tau+\partial_{+}\chi+m_0^2
(\frac{b^2}{4}+\frac{1}{3})z^{2}-\frac{m_0^2}{2}
\left(r^{2}+\frac{1}{9}\left(z_{2}^{2}+z_{3}^{2}\right)\right).
}
The final form of the effective Lagrangian is given by
\begin{align}
L_{\txt{eff}}= & \partial_{+}\vec{y}\partial_{-}\vec{y}+\left[-m_0^2(\frac{b^2}{2}+\frac{2}{3})z^{2}+m_0^2\left(r^{2}+
\frac{1}{9}\left(z_{2}^{2}+z_{3}^{2}\right)\right)\right]\partial_{-}\vec{y}\partial_{-}\vec{y} \notag \\
 & +\partial_{+}\vec{z}\partial_{-}\vec{z}+\left[-m_0^2(\frac{b^2}{2}+\frac{2}{3})z^{2}+m_0^2
\left(r^{2}+\frac{1}{9}\left(z_{2}^{2}+z_{3}^{2}\right)\right)\right]\partial_{-}\vec{z}\partial_{-}\vec{z} \notag \\
 & +\partial_{+}\vec{r}\partial_{-}\vec{r}+\left[-m_0^2(\frac{b^2}{2}+\frac{2}{3})z^{2}+m_0^2
\left(r^{2}+\frac{1}{9}\left(z_{2}^{2}+z_{3}^{2}\right)\right)\right]\partial_{-}\vec{r}\partial_{-}\vec{r}\notag\\
 & -\frac{m_0^2}{9}\frac{z_{2}^{2}+z_{3}^{2}}{4}-m_0^2\frac{r^{2}}{4}.
\end{align}
It is worth to note again that although complicated, this Lagrangian is much simpler than the original one.
It has several advantages. First of all, since it is quite similar to the
reduced model considered in\cite{MS}, it is tempting to expect that it admits a Lax connection although 
it is not clear whether the original theory is integrable. This however can be conclusively
stated only after inclusion of the fermionic part and thorough analysis of the whole model.
Secondly, the analysis of the reduced model 
will be much easier than the original one. It can be performed along the lines of \cite{MS,Zarembo:2007nfs1,Zarembo:2007nfs2,Kluson:2007nfs}.
If the above conjectures are true, 
the S~matrix is expected to be similar to the one conjectured in \cite{MS}. 

\sect{Conclusions}

As we discussed in the Introduction, string backgrounds with less supersymmetry are of great interest,
especially because their gauge theory duals are closely related to QCD. 
Since superstring theory on known backgrounds is highly non-linear and difficult to deal with, 
some limiting procedures proved useful and  allow to extract
more information for the models. One of the limits widely used in the last years is the so-called pp-wave limit. 
It is very attractive because the resulting theory is exactly solvable even on the quantized level. The issues concerning the 
validity of the correspondence have also been studied. Although exactly solvable, some of the important properties and 
structure of the original theory get lost due to the limit. One possibility to weaken the pp-wave limit
and to preserve many of the key features of superstring theory on $\axs$ background
was proposed by Maldacena and Swanson\footnote{See also \cite{gleb:04Sm}.}. 
They called this procedure taking near flat space limit.

Being the string dual of a $\N=1$ gauge theory related to hadron physics, superstring theory on 
the Maldacena-Nunez background is of particular importance.
Inspired by its importance, in this paper we considered the near flat space limit of string theory
on the Maldacena-Nunez background. It is conjectured that, as in the case of string theory on the $\axs$ background,
many of the key features of the original theory should be presented in the reduced model. Our considerations 
show that the reduced model in this background is of the same type as the one obtained from the NFS limit of
$\axs$. The bosonic part of the gauge fixed Lagrangian of the reduced model contain again ``mass terms''. 
The latter are composed differently from  the case considered in \cite{MS}. Based on
the structural similarity to the case in \cite{MS}, we expect that the near flat space limit reduction
of string theory on the Maldacena-Nunez background is integrable. Since in this paper we
consider only bosonic part of the model, one cannot make conclusive statements without
inclusion of the fermionic part and thorough analysis of the resulting dynamical system. 
Certainly further investigations along this line must be done in order to to establish
integrability.
A second conjecture based on our results and the findings
in \cite{BT} is that, as in the case of the pp-wave limit, the near flat space limit reduced models 
have an universal structure - that of \cite{MS}. If true, the techniques of 
\cite{Zarembo:2007nfs1,Zarembo:2007nfs2,pul} will be also universal within the class 
of NFS reduced models. We considered also the softly broken model which in the limit of
small $\rho$ slightly differs from the supersymmetric case. The resulting NFS limit
parameterized by a parameter $b\in(0,2/3]$ has the very same structure and gives further
support to the above conjectures. 

There are several directions for further generalizations of these studies. 
First of all, it would be interesting, to perform detailed analysis
of the integrability of the reduced model. Guided by the close similarity of the structure
of the (bosonic) Lagrangian obtained here and that in \cite{MS}, to look for
a Lax connection for the gauge fixed model and to perform an analysis along the
lines of, say \cite{Kluson:2007nfs}. 
Secondly, one can apply
these techniques to study string sigma models in so-called beta-deformed backgrounds introduced
first in \cite{Lunin:2005jy},
assuming that the NFS limit interpolates between pp-wave and magnon sectors of the theory
\cite{Chu:2006ae,Bobev:2006fg}. The expectations that some key features of the original
model survive the NFS limit must also find answers through further investigations.
One should note that in this study we considered only the bosonic part of the theory.
The near flat limit of the fermionic part is much more difficult due to the presence of non-trivial
RR field, but deserves detailed study which is currently under investigation. 
Although there is some progress in our understanding of the role of the integrable structures
in AdS/CFT, or more generally string/gauge theory duality, much has still to be done.

\bigskip
\leftline{\bf Acknowledgments}
\smallskip

We thank S. Guttenberg for useful discussion at an early stage of the project. 
Many thanks to Gleb Arutyunov for valuable comments, correspondence and
pointing out some references. We appreciate the suggestion by A. Cotrone
to include softly broken case and the useful comments.

R.R. acknowledges the Senior Research Fellowship, the warm hospitality
and stimulating atmosphere at the Erwin Schr\"{o}dinger International
Institute for Mathematical Physics.
The work of C.M. is supported by the Austrian Research Fund FWF grant \# P18679-N16.
R.R. is partly supported by the Bulgarian Research Fund NSF grant BUF-14/06 and  
061/2007 with RF of SU.


\end{document}